# Ground state of the spin-1/2 chain of green dioptase at high fields

Kazuki MATSUI, Masashi FUJISAWA[1] Kenta HAGIWARA, Yukihiro HOSHINO, Takayuki GOTO[*], Takahiko SASAKI[2], Hidekazu TANAKA[3], Susumu Okubo[1] and Hitoshi OHTA[1]

*Physics Division, Sophia Univerisity, Chiyodaku, Tokyo 102-8554, Japan*
*[1] Molecular Photoscience Research Center, Kobe University, Kobe 657-5801, Japan*
*[2] Institute for Materials Research, Tohoku University, Sendai 980-8577, Japan*
*[3] Department of Physics, Tokyo Institute of Technology, Meguro-ku, Tokyo 152-8551, Japan*

*E-mail: gotoo-t@sophia.ac.jp*

The gem-stone dioptase $Cu_6Si_6O_{18}\cdot 6H_2O$ has a chiral crystal structure of equilateral triangular helices consisting of Cu-3*d* spins. It shows an antiferromagnetic order with an easy axis along *c* at $T_N$ = 14.5 K under zero field, and a magnetization jump at $H_C$ = 13.5 T when the field is applied along *c*-axis. By $^{29}$Si-NMR measurements, we have revealed that the high-field state is essentially the two sub-lattice structure, and that the component within *ab*-plane is collinear. The result indicates no apparent match with the geometrical pattern of helical spin chain.

## 1. Introduction

Chirality is a fundamental parameter in nature; those that are under its influence range from molecules to cosmic sciences. Recently, the interplay between crystallographic and magnetic chiralities has been paid attention, because properties of the helimagnetic structure depends on its underlying chiral crystalline structure as represented by MnSi [1-3]. In this article, we focus attention on the gem-stone dioptase $Cu_6Si_6O_{18}\cdot 6H_2O$, which is a transparent green mineral built up from helical spin chains of magnetic $Cu^{2+}$ ions along threefold axis along *c*. These helical chains form a unique threefold spin network, called a dioptase lattice [4]. Our motivation is rooted in a simple interest on whether or not this characteristic crystal structures will affect its magnetic structure [1-5].

The exchange network in dioptase is characterized by two dominant paths $J_c$ and $J_d$, which connect spins within a helix and those in nearest neighboring helices, respectively. Gros *et al.* reported by analyzing the temperature dependence of the magnetic susceptibility that both the two interactions are antiferromagnetic [6]. On the other hand, Janson *et al.* have claimed from the result of quantum Monte Carlo simulations that $J_d$ is ferromagnetic [7]. This disagreement is not settled until now within the knowledge of the authors.

So far, detailed studies on susceptibility, specific heat and neutron diffraction at zero field have reported the existence of long range antiferromagnetic order at $T_N$ = 14.5 K at zero field with the ordered moment of $\mu$ = 0.59 $\mu_B$ and with an easy axis along $c$-axis [8]. While spins on the same level along $c$-axis are aligned ferromagnetically, those on adjacent levels are antiferromagnetic. Recently, it has been reported that an anomalous magnetization jump [9] takes place at $H_C \cong 13$ (±1) T when $H//c$, and that the magnetization above $H_C$ does not coincides with that for $H \perp c$. This suggests that a high-field spin state cannot be explained by the conventional spin-flop. Ohta *et al.* have also reported by the high-field ESR [10,11] up to 55 T that the frequency of antiferromagnetic resonances (AFMR) above 12.5 T shows a significant deviation from the theory of conventional two-sublattice model [12].

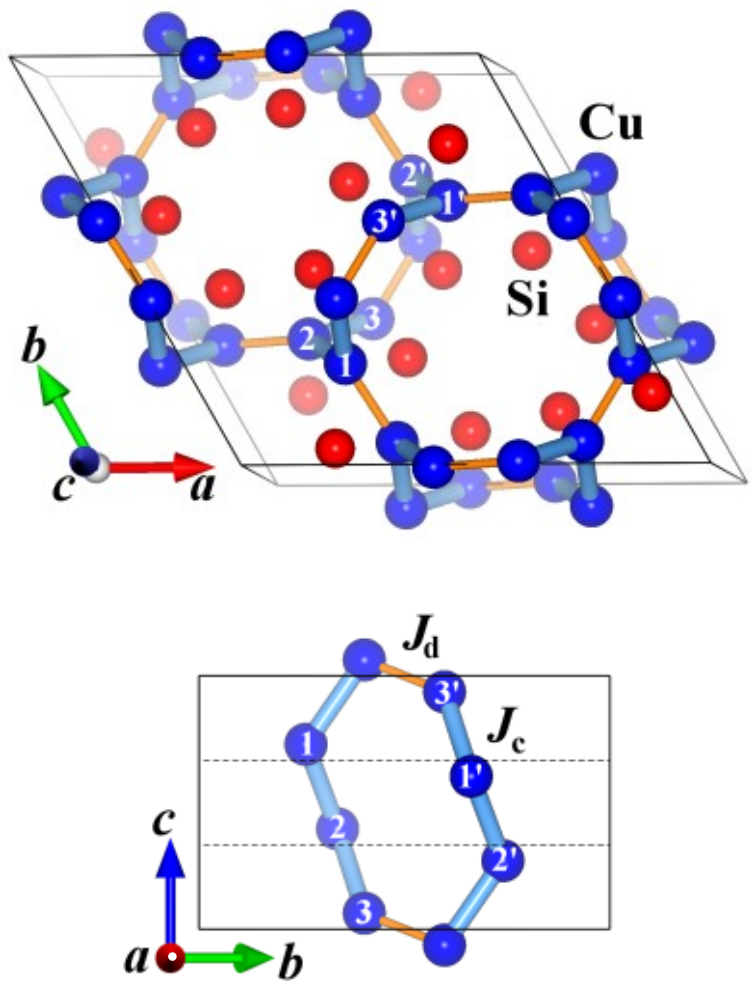

**Fig. 1.** (Color online) Schematic drawings of the location of Cu and Si atoms within *ab* plane (top) and *bc* plane (bottom). The solid-line boxes show an unit cell. The numbers and prime symbols on atoms shows the position along $c$-axis.

In order to investigate the spin structure at high fields above $H_C$, we have measured $^{29}$Si-NMR spectra under a wide range of fields up to 18 T. Its peak-splitting pattern in the ordered state showed a drastic change with increasing field across $H_C$. We have succeeded to reproduce this change in the split pattern by calculating the local field at the Si site based on the simple two-sublattice model. The result shows that the underlying helical crystal structure does not influence the magnetic structure in the present compound, and hence urges the re-investigation or re-interpretation of other physical quantities.

## 2. Experimental

The single crystal of the natural mineral green dioptase is used for measurements of NMR and the specific heat. Note that the dehydrated black dioptase has a different crystal structure and hence magnetic properties[13]. The volume of the crystal is approximately 150 mm$^3$ with hexagonal-rod shape. NMR spectra were obtained by recording the amplitude of spin-echo signal against the applied field up to 18 T [14]. The specific heat under magnetic field up to 9 T was measured by a commercial multipurpose device (PPMS, Quantum Design). A distinct lambda-type peak was observed in its temperature dependence, and the Néel Temperature $T_N$ was determined as the top of peak.

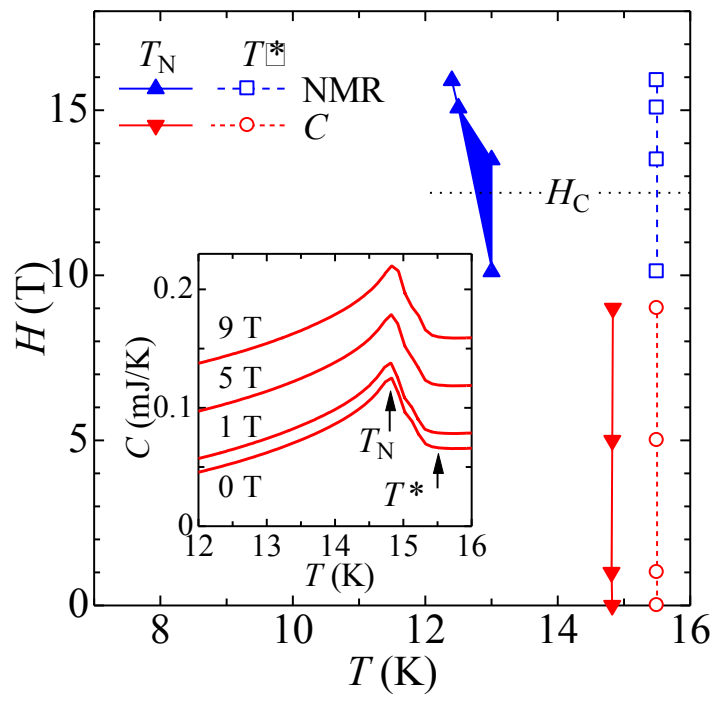

**Fig. 2.** (Color online) The field dependence of the Néel temperature $T_N$ and the onset temperature $T$* determined by the specific heat $C$ and NMR. The inset shows the temperature dependence of $C$ with the definitions of $T_N$ and $T$*.

Before describing NMR results, we refer in detail the local crystal structure that determines the hyperfine field at the Si site. Fig. 1 shows the schematic drawing of Cu and Si location. Cu atoms compose equilateral-triangular helices along *c*-axis. There are two types of helices in a unit cell, that is, the right-handed and left-handed screws; they are located alternately within *ab*-plane, forming an equilateral dodecagon. The neighboring two dodecagons share two helices connected with $J_d$ exchange interaction. The direction of this $J_d$ bond is slightly out of alignment with *a* or *b* axes. The Si atoms are in the midst of $SiO_4$ tetrahedron that forms a hexagon within *ab*-plane, and touches internally the surrounding dodecagon. The distance between the nearest neighboring Cu and Si is approximately 3.09 Å [4], which is near enough to expect to bring an appreciable hyperfine field at Si-NMR.

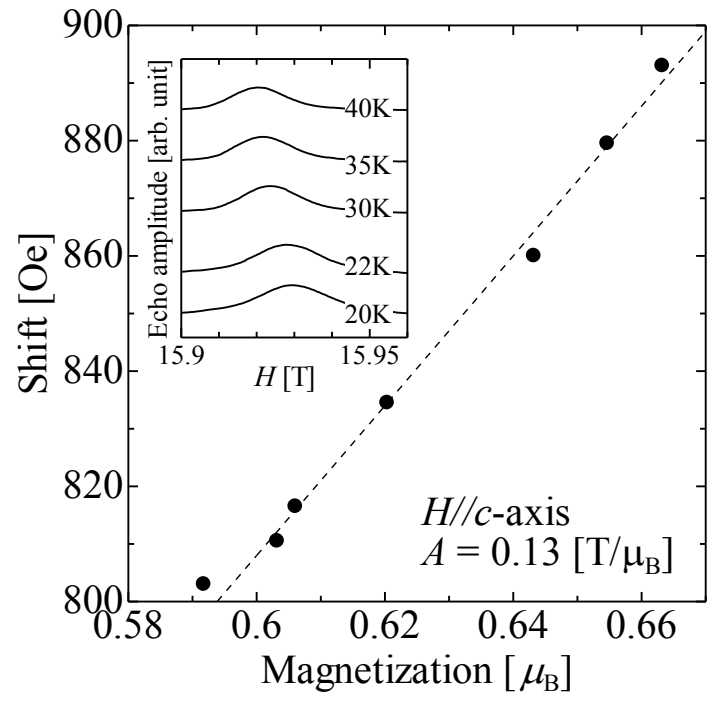

**Fig. 3.** The plot of $^{29}$Si-NMR shift versus uniform magnetization [2] with an implicit parameter of temperature between 20 and 40 K. The inset shows typical profile spectra in the paramagnetic state.

## 3. Results

Figure 2 shows the temperature dependence of the specific heat $C$ under various magnetic fields. The two characteristic temperatures $T_N$ and $T^*$ were determined as a peak position of the lambda peak, and as the onset of the increase in $C$ with lowering temperature, respectively. Both $T_N$ and $T^*$ were field independent up to 9 T.

Next, in Fig. 3 we show the plot of the NMR shift versus uniform magnetization with temperature as an implicit parameter in the paramagnetic state. From the gradient of linear dependence between the shift and the uniform magnetization [14], we have obtained the transferred hyperfine coupling constant at the Si site as $A$ = 0.13 T/$\mu_B$. This moderately large value allows us to probe the Cu 3*d*-spin state by Si-NMR.

Figure 4 shows typical profiles of spectra above and below $T_N$, and the temperature dependence of the hyperfine field $\Delta H$ at the Si site, determined as the half of the peak-splitting width. The peak splitting clearly demonstrates the existence of the simple antiferromagnetic order. As lowering temperature, $\Delta H$ first shows a marginal rise at $T^*$, which coincides with the onset temperature of $C$, and starts increasing steeply at still lower temperature $T_N$, which is approximately 2.5 K lower than $T^*$. The $T_N$ determined thus by NMR is approximately 1 K lower than that determined by the specific heat.

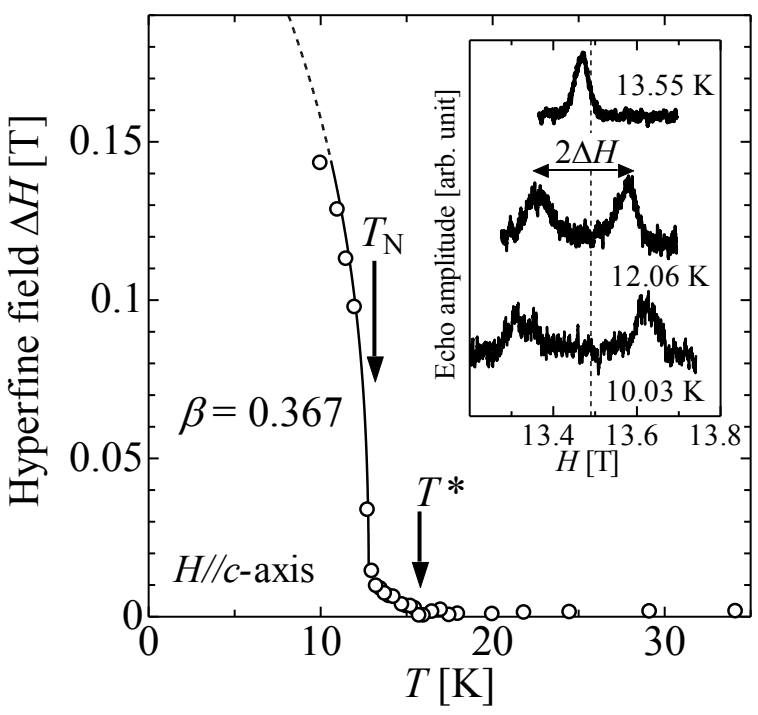

**Fig. 4.** The temperature dependence of the hyperfine field $\Delta H$ at the Si site. The solid curve shows the function $|T_N - T|^{\beta}$. The inset shows typical spectra above and below $T_N$.

The onset temperature $T^*$ may correspond to a precursor of the long range antiferromagnetic order or in other words the short-range order above $T_N$. The temperature dependence of $\Delta H$ below $T_N$ was well described by the power law function $|T_N - T|^{\beta}$ with the critical exponent of $\beta = 0.367$, which agrees with the 3D-Heisenberg model [15,16].

The field region of NMR is approximately 13.2 – 13.7 T, which well falls within the field region of broad magnetization jump at $H_C$ = 13 (±1) T [9]. So, we consider that the result shown in Fig. 4 represents the low-field spin state below $H_C$.

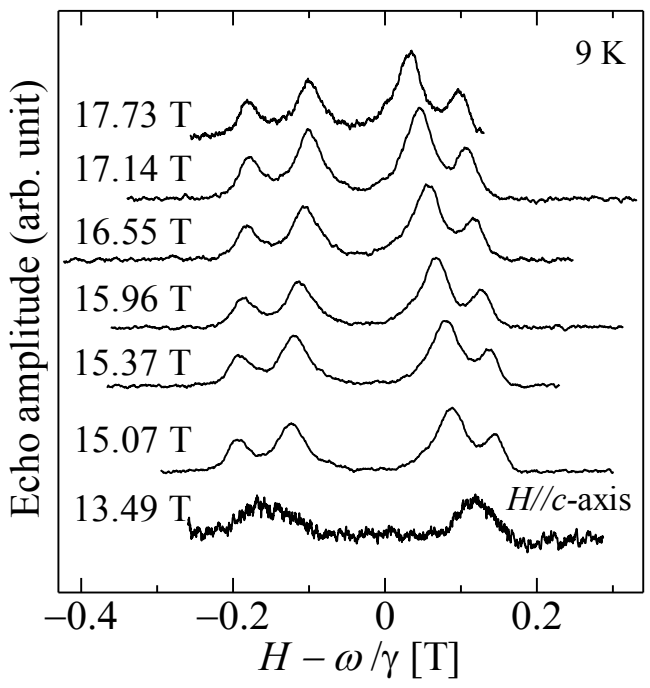

**Fig. 5.** Typical profile of NMR spectra at and above $H_C$ = 13.5 T.

At higher field above $H_C$, the splitting pattern drastically changes. There appeared four peaks, located symmetrically around the zero shift position as shown in Fig. 5. The amplitude of the inner two peaks is approximately two times of those outer two. The distance between the left side pair and the right side pair does not change as passing $H_C$. With still increasing the field, it showed a slight reduction.

## 3. Discussion

The drastic change in the splitting pattern of NMR spectra as passing $H_C$ clearly demonstrate the existence of spin flop transition as reported by magnetization and ESR measurements [9,10]. In order to determine the spin structure above $H_C$, we have calculated the hyperfine field at each Si site in a unit cell by adding up the contribution from surrounding Cu-3$d$ spins within 3×3×3 unit cells [17]. The interaction between electron and nuclear spins is assumed to be the classical dipole type. We extracted from the sum the parallel component along the applied field to obtain the hyperfine field [17].

Before proceeding to the high-field spin structure, we first confirmed the low-field antiferromagnetic structure, that is, $\mu$ = 0.59 $\mu_B$, aligned along $c$-axis and ferromagnetic within $ab$-plane as reported by neutron diffractions

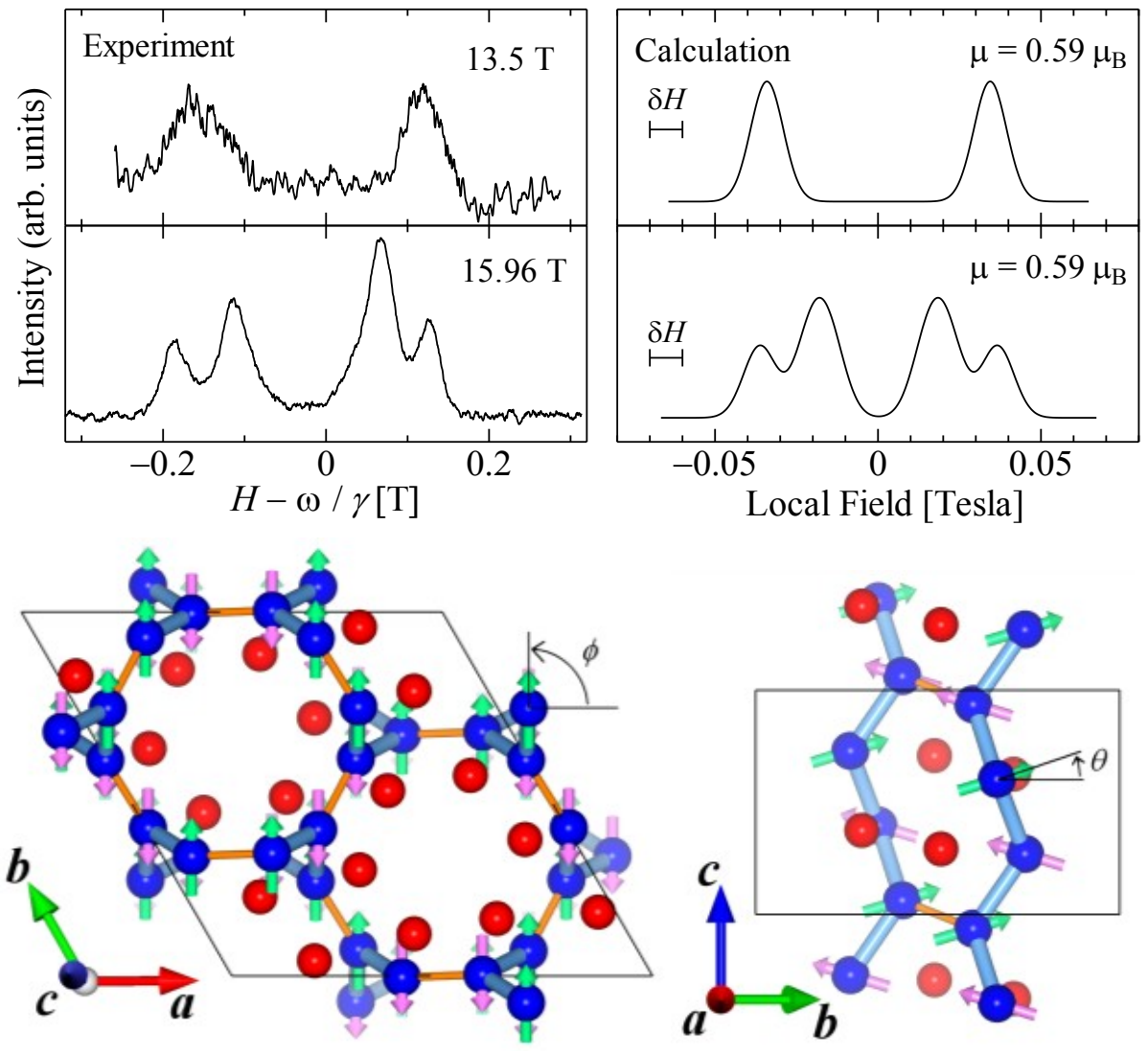

**Fig. 6.** (Color online) (top) Observed spectra and calculated one below and above $H_C$. (below) The schematic drawings of the spin structure model at high field $H > H_C$ with the view along $c$-axis (left) and $b$-axis (right), respectively. The angles $\theta = 2.87°$ and $\phi = 84°$ show the direction of the ordered moment.

[8]. The calculation confirmed the appearance of two hyperfine fields at Si sites in a unit cell. We plot the obtained hyperfine fields with a Gaussian convolution of 100 Oe in Fig. 6 to compare with observed NMR spectrum.

The calculation well reproduces the experimentally obtained spectrum, though the absolute value of the calculated splitting width is a little bit smaller than the observation. There are two possible reasons; one is the case that the actual length of ordered moments is larger than was reported by neutron experiments [8], and the other is due to the possible contribution of supertransferred hyperfine interaction via ligand oxygen atoms intervening Si and Cu.

Next, we raise the field above $H_C$. There of course are vast possible models, so in choosing spin-structure candidates, we put a constraint that the uniform magnetization that arises from a cant toward *c* matches the reported value under the applied field. The tested models include the classical 120° structure with chirality $\chi = \pm 1$ and with a uniform cant, the umbrella state with asymmetric cant and that associated with the conventional spinflop with uniform cant along the applied field. Among these candidates, the last one shown schematically in Fig. 6 was found to be most plausible to reproduce the observed spectra under $H > H_C$. The phase difference in the spin alternation between the two neighboring helices is very important. The observed spectrum is reproduced only when the two spins connected by $J_d$ are set ferromagnetic. Otherwise, the hyperfine field at the Si site accidentally cancels out to be extremely small. This observation favors the Janson's report on the sign of exchange interaction parameter, that is $J_d < 0$ [7].

The cant angle from the *ab*-plane $\theta = 2.87°$ was chosen so that $\cos\theta = M(H)/\mu$, where $M(H)$ is the uniform magnetization under the applied field $H$, and $\mu = 0.59\ \mu_B$ is the ordered moment [8]. The azimuth angle $\phi$ within *ab*-plane was found to be approximately 84°. With this $\phi$, two values of hyperfine fields accidentally coincide and hence produce the difference in the peak height of inner and outer pairs as shown in Fig. 6. The slight deviation of $\phi$ from the normal angle comes from the fact that the direction $J_d$ bond is out of alignment from *a* or *b*-axes as stated above.

Due to the rhombohedral crystal symmetry, there exist three equivalent values for $\phi$, that is, 84°, 204° and 324°. Although this suggests the existence of a domain structure, it is not self-evident whether or not it actually exists, because within $J_c$-$J_d$ approximation, there is no frustration effect that competes with the antiferromagnetic spin structure in this system. In order to confirm the possibility of the domain structure, an NMR measurement under the field slightly tilted from *c*-axis is necessary, which is now in preparation.

## 5. Summary

We have investigated the spin structure of the dioptase at high fields by $^{29}$Si-NMR. The observed four-peak type spectrum at high field above $H_C$ was well reproduced by the simulation based on the two sub-lattice spin structure with a slight and uniform cant toward *c*-axis. The in-plane component of the spin direction is collinear and nearly perpendicular to *a*-axis. This indicates that in the present system the spin structure have less effect from the geometrical pattern of exchange paths. The neighboring two spins on two adjacent helices were found to direct ferromagnetically, indicating that the exchange interaction $J_d$ is ferromagnetic.

## Acknowledgment

This work was partially supported by JSPS KAKENHI Grant Number 21110518, 24540350 and 21540344. The NMR measurements were performed in part at High Field Laboratory for Superconducting Materials, Institute for Materials Research, Tohoku University.

## References

[1] V Dmitriev *et al.*: J. Phys. Condens. Matt. **24** (2012) 366005.
[2] R. D. Johnson *et al.*: Phys. Rev. Lett. **111** (2013) 017202.
[3] Y. Kousaka *et al.*: J. Phys. Soc. Jpn. **76** (2007) 123709.
[4] P. H. Ribbe *et al.*: Am. Mineral. 62 (1977) 807.
[5] S. Nakajima *et al.*: J. Phys. Soc. Jpn. **81** (2012) 063706.
[6] C. Gros *et al.* Europhys. Lett., **60** (2), pp. 276–280 (2002).
[7] O. Janson *et al.*: Phys. Rev. B82 (2010) 014424.
[8] E. L. Belokoneva *et al.*: Phys Chem Minerals (2002) 29, 430.
[9] H. Ohta *et al*. J. Phys.: Conf. Ser. **150** (2009) 042151.
[10] S. Ohkubo *et al.*: Phys. Rev. B86 (2012) 140401(R).
[11] M. Fujisawa *et al.*: Phys. Rev. B80 (2009) 012408.
[12] H. Ohta *et al.*: Appl. Magn. Reson. 35 (2009) 399.
[13] M. Baenitz *et al.*: Phys. Rev. 62 (2000) 12201.
[14] H. Inoue *et al.*: Phys. Rev. B79 (2009) 174418.
[15] T. Saito *et al.*: Phys. Rev. B74 (2006) 134423.
[16] K. Kanada *et al.*: J. Phys. Soc. Jpn. 76 (2007) 064706.
[17] K. Kanada *et al.*: J. Phys. Chem. Solids 68 (2007) 2191–2194.